\documentclass{ws-mpla}
\newcommand{\half}{\frac{1}{2}}
\newcommand{\thrd}{\frac{1}{3}}
\newcommand{\frth}{\frac{1}{4}}
\newcommand{\sxth}{\frac{1}{6}}
\begin{document}
\markboth{R. K. Nesbet}{Cosmological implications of conformal field 
theory}
\catchline{}{}{}{}{}
\title{COSMOLOGICAL IMPLICATIONS OF CONFORMAL FIELD THEORY}
\author{\footnotesize ROBERT K. NESBET}
\address{IBM Almaden Research Center\\
650 Harry Road, San Jose, CA 95120, USA\\
rkn@almaden.ibm.com}
\maketitle
\begin{history}
\received{(received date)}
\revised{(revised date)}
\end{history}
\begin{abstract}
Requiring all massless elementary fields to have conformal scaling 
symmetry removes a conflict between gravitational theory and the quantum
theory of elementary particles and fields.  
Extending this postulate to the scalar field of the Higgs model, 
dynamical breaking of both gauge and conformal symmetries determines 
parameters for the interacting fields.  In uniform isotropic geometry 
a modified Friedmann cosmic evolution equation is derived with
nonvanishing cosmological constant.  
Parameters determined by numerical solution are consistent with
empirical data for redshifts $z\leq z_*=1090$, including luminosity
distances for observed type Ia supernovae and peak structure ratios in
the cosmic microwave background (CMB). The theory does not require dark
matter.
\end{abstract}
\ccode{PACS Nos.: 04.20.Cv, 98.80.-k, 11.15.-q}

\section{Introduction}
\par Massless fields of the standard model\cite{PAS95,CAG98} have
definite conformal character for local Weyl scaling,\cite{MAN06,DEW64}
such that $g_{\mu\nu}(x)\to g_{\mu\nu}(x)e^{2\alpha(x)}$. If an action 
integral is conformally invariant, the implied energy-momentum 4-tensor 
is traceless.  The Einstein tensor is not.  Compatibility can be imposed
if Einstein-Hilbert gravitational field action is replaced by a uniquely
determined action integral $I_g$ constructed using the conformal
Weyl tensor.\cite{MAN06}  This preserves phenomenology on the distance
scale of the solar system, while providing an alternative explanation
of excessive rotational velocities in galaxies, without invoking dark
matter.\cite{MAN06}
\par Higgs symmetry-breaking invokes a spacetime constant scalar
field $\Phi$.  Conformal symmetry requires Lagrangian density  
${\cal L}_\Phi$ to contain a term proportional to
Ricci curvature scalar $R=g_{\mu\nu}R^{\mu\nu}$, where
Ricci tensor $R^{\mu\nu}$ is the symmetric contraction 
$R_\lambda^{\mu\lambda\nu}$ of the gravitational Riemann tensor.
Thus conformal $\Phi$ is a cosmological entity.\cite{MAN06}  
Its energy-momentum tensor contains a residual 
cosmological constant.\cite{MAN06}   
An earlier study\cite{CCJ70} showed that a conformal scalar field can
modify Einstein-Hilbert gravitation, but did not consider the 
conformally invariant Weyl tensor.
\par These results suggest a unifying postulate, that all massless
elementary fields have conformal symmetry.  The cosmological
consequences of universal conformal symmetry are explored here.
It will be shown that identifying the Higgs scalar field as the  
source of a cosmological constant (aka dark energy) produces an 
internally consistent theory, in agreement with all relevant 
cosmological data. 
\par Variational formalism of classical field theory\cite{NES03} is
easily extended to the context of general relativity.\cite{MAN06}
Generic Lagrangian density ${\cal L}$ defines action integral
$I=\int d^4x\sqrt{-g}{\cal L}$, where $g$ is the determinant of metric 
tensor $ g_{\mu\nu}$.  The metric functional derivative
$\frac{1}{\sqrt{-g}}\frac{\delta I}{\delta g_{\mu\nu}}$ is
$X^{\mu\nu}=x^{\mu\nu}+\half{\cal L}g^{\mu\nu}$, 
if $\delta{\cal L}=x^{\mu\nu}\delta g_{\mu\nu}$.
This defines symmetric energy-momentum tensor 
$\Theta^{\mu\nu}=-2X^{\mu\nu}$.  Evaluated for solutions of the
conformal field equations, trace $g_{\mu\nu}\Theta^{\mu\nu}=0$.
\par Conformally invariant action integral $I_\Phi$ is defined for
scalar field $\Phi$ by Lagrangian density 
${\cal L}_\Phi=
 (\partial_\mu\Phi)^\dag\partial^\mu\Phi-\sxth R\Phi^\dag\Phi
 -\lambda(\Phi^\dag\Phi)^2$,
where $R$ is the Ricci scalar.\cite{MAN06}  The Higgs  
mechanism\cite{CAG98} postulates incremental Lagrangian density 
$\Delta{\cal L}_\Phi=w^2\Phi^\dag\Phi-\lambda(\Phi^\dag\Phi)^2$.
This augments ${\cal L}_\Phi$
by a term $w^2\Phi^\dag\Phi$ which breaks conformal symmetry. 
In conformal theory, this term must be produced dynamically.  
\par Mannheim\cite{MAN06}[arXiv:astro-ph/0505266] has recently 
reviewed conformal gravitational theory.
The Einstein-Hilbert Lagrangian density is replaced by a 
uniquely determined quadratic form,\cite{MAN06}[Section 8.7]
${\cal L}_g=-\alpha_gC^\lambda_{\mu\kappa\nu}C_\lambda^{\mu\kappa\nu}$,
where $C^\lambda_{\mu\kappa\nu}$ is the conformally invariant Weyl
tensor.  This tensor is the traceless component of Riemann tensor
$R^\lambda_{\mu\kappa\nu}$, obtained by removing a linear combination
of contracted terms depending on the Ricci tensor and 
scalar,\cite{MAN06}[Eq.(180)]. The Weyl tensor vanishes identically in
uniform, isotropic Robertson-Walker (RW) geometry.  Vanishing of the  
metric functional derivative of action integral $I_g$,\cite{MAN06}
[Eq.(185)], for the RW metric given below, can be verified by
direct evaluation. 
\par  Thus the conformal gravitational action integral replaces the 
standard Einstein-Hilbert action integral, but in the uniform
model of cosmology its functional derivative drops out completely from
the gravitational field equations,\cite{MAN06}[Section 10.1].
The observed Hubble expansion requires an alternative gravitational
mechanism.  This is supplied by a postulated conformal scalar field. 
A nonvanishing conformal scalar field determines gravitational field
equations that differ from Einstein-Hilbert theory.  
The Newton-Einstein gravitational constant is not relevant.  
As shown by Mannheim,\cite{MAN06}[Eq.(224)], 
the gravitational constant determined by the scalar field is inherently
negative, appropriate to Hubble expansion of the early universe.
\par
The argument here differs from Mannheim by noting that the Lagrangian 
terms proportional to $\Phi^\dag\Phi$ in Higgs and conformal theory
have opposite algebraic signs.  A consistent theory must include both.  
The consequences of this are examined here, leading in particular to
a modified Friedmann cosmic evolution equation that differs from the
standard form used in all previous work.  An important consequence is
that the spatial curvature parameter implied by the modified Friedmann
equation is now consistent with current cosmological data, removing
a severe problem in fitting type Ia supernovae redshift data 
using the standard equation.\cite{MAN06} 
The anomalous imaginary-mass term in the Higgs scalar field Lagrangian
becomes a cosmological constant (dark energy) in the modified Friedmann
equation.  This term dominates the current epoch.
\par It should be noted that fitting conformal gravitation to 
galactic rotation data\cite{MAN06}[Section 9.3] implies a universal 
nonclassical linear gravitational potential $V=\gamma_0c^2r/2$.
Coefficient $\gamma_0$, independent of galactic luminous mass, must
be attributed to the background Hubble flow.\cite{MAN06} 
On converting the local Schwarzchild metric to conformal RW form, this
produces a curvature parameter $k=-\frth\gamma_0^2$ which is small and
negative, consistent with other current empirical data.  This supports 
the present argument for modifying the standard Friedmann equation 
in RW geometry.
\par For redshift $z(t)$, the modified Friedmann equation determines
scale parameter $a(t)=1/(1+z(t))$ and Hubble function
$H(t)=\frac{{\dot a}}{a}(t)$.  Acceleration parameter 
${\ddot a}a/{\dot a}^2$ is always positive and occurs explicitly
in the modified dimensionless sum rule.  In the current epoch,
dark energy and acceleration terms are of comparable magnitude, 
the curvature term is small, and other terms are negligible. 
As $t\to\infty$, $H(t)$ descends asymptotically to a finite
value determined by the cosmological constant, while the acceleration
parameter goes to zero. 
\par The present analysis derives Ricci scalar $R$ as a 
time-dependent function.  This indicates that other
nominally constant parameters deduced here are time-dependent
on a cosmological scale (ten billion years).  
If these parameters were strictly constants, fitted here to data for
$z\leq z_*=1090$, the Hubble function would increase from zero at some
initial minimum $a_0$ ($z_0>z_*$) to a maximum value ${\bar H}$ at
${\bar a}<a_*$, defining an inflationary epoch.  Comoving radius $1/aH$ 
would decrease monotonically from infinity, which eliminates any
inherent horizon problem.\cite{DOD03}  Details of this early epoch
require accurate time-dependent parameters, not currently available.
An initial big-bang singularity, or initial $a_0$ small enough that the
corresponding temperature would support nucleosynthesis,\cite{DOD03}
cannot be ruled out.

\section{Scalar and tensor field equations}
\par The scalar field equation is 
$\partial_\mu\partial^\mu\Phi= 
 (-\sxth R+w^2-2\lambda\Phi^\dag\Phi)\Phi$.
Generalizing the Higgs construction, for constant $R$ the scalar field
equation has a global solution\cite{NES08} such that 
$\Phi^\dag\Phi=\phi_0^2=(w^2-\sxth R)/2\lambda$, if this ratio is 
positive.  $\phi_0^2$ determines gauge boson masses. 
Empirical parameters imply $\sxth R>w^2$, so that $\lambda<0$.
Although this differs from the standard electroweak model, which 
assumes positive $w^2$ and $\lambda$, conformal theory determines a
stable scalar field solution with a finite energy-momentum tensor
so long as $\phi_0^2>0$.\cite{NES10b} 
\par If Ricci $R$ were neglected, for positive parameter $\lambda$ 
fluctuations about scalar field $\Phi^\dag\Phi=\phi_0^2$ would 
satisfy a Klein-Gordon equation with mass parameter $m_H=\sqrt{2w^2}$,
defining a Higgs boson.\cite{CAG98}  The empirical value of $w$ deduced
here would imply $m_H\simeq 10^{-33}eV$.  However, the implied value
of $\lambda$ is negative, inconsistent with a Klein-Gordon equation.
This does not define a Higgs mass.  This issue is discussed in more 
detail elsewhere.\cite{NES10b}
\par Given matter/radiation action integral $I_m$, the gravitational
field equation in terms of functional derivatives is 
$X^{\mu\nu}_g+X^{\mu\nu}_\Phi+X^{\mu\nu}_m$=0.  Gravitational
energy-momentum exactly cancels that of matter and radiation. Conformal 
$X_g$ vanishes identically in uniform, isotropic geometry.\cite{MAN06}
RW gravitation is determined by $X_\Phi$, due to Ricci
scalar $R$ in ${\cal L}_\Phi$.  Finite $X_m$ determines finite $X_\Phi$.
For the quantized scalar field, finite energy density precludes 
destabilization of the vacuum state.\cite{NES10b} 
The resulting gravitational field equation is 
$X_\Phi^{\mu\nu}=\half\Theta_m^{\mu\nu}$, where $X_\Phi^{\mu\nu}=
\sxth R^{\mu\nu}\Phi^\dag\Phi+\half{\cal L}_\Phi g^{\mu\nu}$,
evaluated for $\phi_0^2=(w^2-\sxth R)/2\lambda$. 
\par For ${\cal L}_\Phi=\phi_0^2(w^2-\sxth R-\lambda\phi_0^2)
=\sxth\phi_0^2(3w^2-\half R)$,
the gravitational functional derivative is
$X_\Phi^{\mu\nu}
=\sxth\phi_0^2(R^{\mu\nu}-\frth Rg^{\mu\nu}+\frac{3}{2}w^2g^{\mu\nu})$.
The first two terms here replace the Einstein tensor of standard theory
by a traceless modified tensor.  This removes an obvious
inconsistency in the context of universal conformal symmetry.
Defining ${\bar\kappa}=-3/\phi_0^2$ and ${\bar\Lambda}=\frac{3}{2}w^2$,
the gravitational field equation is 
$R^{\mu\nu}-\frth Rg^{\mu\nu}+{\bar\Lambda}g^{\mu\nu}
 =-{\bar\kappa}\Theta_m^{\mu\nu}$.  
The effective cosmological constant ${\bar\Lambda}$ here is identified 
with the Higgs scalar field parameter $w^2$.

\section{The modified Friedmann equation}
\par For Robertson-Walker metric 
$ds^2=dt^2
-a(t)^2(\frac{dr^2}{1-kr^2}+r^2d\theta^2+r^2\sin^2\theta d\phi^2)$,
Ricci tensor $R^{\mu\nu}$ depends on $a(t)$ through two independent 
functions, 
$\xi_0(t)=\frac{\ddot a}{a}$ and
$\xi_1(t)=\frac{{\dot a}^2}{a^2}+\frac{k}{a^2}$,
such that $R^{00}=3\xi_0$ and scalar $R=6(\xi_0+\xi_1)$.
\par In a consistent conformal theory, vanishing trace eliminates one
of the two independent Friedmann equations of standard theory.
Energy density $\rho=\Theta_m^{00}$ implies modified Friedmann equation
$-\frac{2}{3}(R^{00}-\frth R)= \xi_1(t)-\xi_0(t)= 
\frac{{\dot a}^2}{a^2}+\frac{k}{a^2}-\frac{\ddot a}{a}
    =\frac{2}{3}({\bar\kappa}\rho+{\bar\Lambda})$, which determines 
RW scale parameter $a(t)$ and Hubble function $H(t)$.\cite{NES08}  For
consistency with electroweak theory ${\bar\Lambda}=\frac{3}{2}w^2>0$.  
For positive energy density $\rho$, ${\bar\kappa}\rho$ is negative, 
as shown by Mannheim.\cite{MAN03}
\par At present time $t_0$, $a(t_0)=1$. $H(t_0)=1$ in Hubble units
$H_0=100h$km/s/Mpc, with $h=0.705$\cite{KOM09} and $\hbar,c=1$.
Scaled energy densities $\rho_m a^3$ and $\rho_r a^4$, for matter
and radiation respectively, are constant. In the absence of dark matter,
$\rho_m\simeq\rho_b$, the baryon density.
\par It is convenient to define constant parameters
$\alpha=\frac{2}{3}{\bar\Lambda}=w^2>0$,
$k\simeq0$, $\beta=-\frac{2}{3}{\bar\kappa}\rho_m a^3>0$,
and if $\rho_m\to\rho_b$, $\gamma=3\beta/4R_b(t_0)>0$, where
$\frac{4}{3}R_b(t)=\frac{\Omega_b}{\Omega_\gamma}a(t)$ is the ratio
of baryon to radiation energy densities.
Empirical value\cite{KOM09} $R_b(t_0)=688.6$ is assumed here.
The equation to be integrated is
$\frac{{\dot a}^2}{a^2}-\frac{{\ddot a}}{a}=
 -\frac{d}{dt}\frac{{\dot a}}{a}={\hat\alpha}=
 \alpha-\frac{k}{a^2}-\frac{\beta}{a^3}-\frac{\gamma}{a^4}$.
Dividing by $H^2(t)$ implies dimensionless sum rule
$\Omega_m(t)+\Omega_r(t)+\Omega_\Lambda(t)+\Omega_k(t)+\Omega_q(t)=1$,
where
$\Omega_m(t)= \frac{2}{3}\frac{{\bar\kappa}\rho_m(t)}{H^2(t)}<0$,
$\Omega_r(t)= \frac{2}{3}\frac{{\bar\kappa}\rho_r(t)}{H^2(t)}<0$,
$\Omega_\Lambda(t)=\frac{w^2}{H^2(t)}>0$,
$\Omega_k(t)=-\frac{k}{a^2(t)H^2(t)}$, and
$\Omega_q(t)=\frac{{\ddot a}a}{{\dot a}^2}=-q(t)$.
Acceleration parameter $\Omega_q(t)$ appears explicitly in the sum rule.
\par Because parameters $\alpha, \beta, \gamma$ are all necessarily
positive, ${\hat\alpha}$ must vanish for some  $a({\bar t})={\bar a}$,
defining a maximum value $\frac{{\dot a}}{a}=H({\bar t})={\bar H}$.
Assuming constant parameters, $H(t)=0$ for a finite minimum $a_0$,
defining time $t=0$.  There are no mathematical singularities.
$a(t)$ increases monotonically from $a_0$ to $a({\bar t})={\bar a}$,
where ${\hat\alpha}$ vanishes and Hubble function $H(t)=
\frac{{\dot a}}{a}$ rises to its maximum value ${\bar H}$.
Acceleration parameter
$\Omega_q=1-\frac{{\hat\alpha}}{H^2}=1+\frac{\gamma}{a^4H^2}
(1+\frac{\beta a}{\gamma}+\frac{ka^2}{\gamma}
-\frac{\alpha  a^4}{\gamma})\to\infty$ at $a_0$.
It decreases monotonically to unity at ${\bar t}$, and asymptotically
to zero, while $\Omega_{\Lambda}\to 1$.
\par In standard theory an initial inflationary epoch is postulated, 
in which the comoving Hubble radius $\frac{1}{aH}$ decreases as time
increases, for positive $\Omega_q$.\cite{DOD03}
This condition is satisfied by the modified Friedmann equation.
For $H(t)$ in units of $H_0$,
$\frac{d}{dt}\ln\frac{1}{aH}=
-\frac{{\dot a}}{a}-\frac{{\dot H}}{H}=-H(1+\frac{{\dot H}}{H^2})=
-H(1+\frac{1}{H^2}(\frac{{\ddot a}}{a}-H^2))=
-H\frac{{\ddot a}a}{{\dot a}^2}=-H\Omega_q\leq 0$. 
Scaled Hubble radius $\frac{1}{aH}$ 
decreases monotonically to present value unity.
\par For an integrated distance such as $r_a=\int cdt/a(t)$,
the corresponding geodesic distance in curved space is denoted here
by $d_a=\frac{\sinh(\sqrt{-k}r_a)}{\sqrt{-k}}$ (for $k<0$).
Integral $r_s=\int_0^{t_0}\frac{cdt}{a}=
\int_{a_0}^1\frac{da}{a}\frac{c}{aH}$ defines comoving horizon
$\eta_0=d_s$,\cite{DOD03} monotonically increasing with $t_0$.

\section{Numerical solution in the current epoch}
\par The modified Friedmann equation has been solved for vanishing $k$,
$\beta$, and $\gamma$. Parameter $\alpha$ is adjusted to match a fit 
to observed magnitudes of type Ia supernovae,\cite{MAN03} using scaled 
luminosity distance $H_0d_L/c$ computed as a function of redshift $z$.
$\Omega_q$ is determined by the modified equation.
\begin{table}[h]
\begin{tabular}{lcccc}
z& $\Omega_\Lambda$&$\Omega_q$&$H_0d_L/c(calc)$
 &$H_0d_L/c$(\cite{MAN03})\\ \hline
0.000& 0.732& 0.268& 0.000& 0.000\\
0.063& 0.672& 0.328& 0.066& 0.066\\
0.133& 0.619& 0.381& 0.145& 0.145\\
0.211& 0.571& 0.429& 0.240& 0.241\\
0.298& 0.530& 0.470& 0.355& 0.357\\
0.395& 0.492& 0.508& 0.494& 0.497\\
0.503& 0.459& 0.541& 0.663& 0.666\\
0.623& 0.428& 0.572& 0.868& 0.871\\
0.758& 0.401& 0.599& 1.118& 1.121\\
0.909& 0.376& 0.624& 1.424& 1.426\\
1.079& 0.353& 0.647& 1.799& 1.799\\
\end{tabular}
\end{table}
\par 
Results for $z\leq 1$ agree to graphical accuracy
with the $\Lambda CDM$ model.\cite{MAN03,MAN06}
Parameter $\alpha=\Omega_\Lambda(t_0)=0.732$ for $\Omega_k(t)=0$
is consistent with current empirical values
$\Omega_\Lambda=0.726\pm0.015, \Omega_k=-0.005\pm 0.013$.\cite{KOM09}
Any significant discrepancy would invalidate the present theory.
Mannheim fits type Ia supernovae luminosities setting $\Omega_m=0$
and using the standard Friedmann equation, which requires 
$\Omega_k=1-\Omega_\Lambda$.  The implied $\Omega_k$ is much larger 
than empirical limits $\pm\Omega_k\simeq0.01$.  This is corrected 
by modified sum rule $\Omega_k=1-\Omega_\Lambda-\Omega_q\simeq 0$.
Here $d_L(z)=(1+z)d_\chi(z)$ for geodesic distance $d_\chi$ such 
that $r_\chi=\chi(z)=\int_{t_z}^{t_0}cdt/a(t)$, for $z(t_z)=z$.
Parameters $\Omega_m, \Omega_r$, which scale as $a^{-3}, a^{-4}$
respectively, can apparently be neglected for $z\leq 1$.

\section{Recombination epoch}
\par In standard theory 
$aH|_{a\to 0}\to(\Omega_mH_0^2)^\half(1+z)^\half$
is used to define dimensionless shift ratio\cite{WAM07,KOM09} 
$R(z)= (\Omega_m H_0^2)^\half(1+z)d_A(z)=aH(1+z)^\half d_A(z)$,
for angular diameter distance $d_A=d_L/(1+z)^2$. 
Not restricted to a particular limiting form, 
$R(z)=\frac{H_0d_A(z)}{c(1+z)^{1/2}}\frac{H(z)}{H_0}
 =\frac{H_0d_L(z)}{c(1+z)^{5/2}}\frac{H(z)}{H_0}$.
The empirical value for recombination epoch
$z_*=1090$, at $t=t_*$, is $R(z_*)=1.710 \pm 0.019$.\cite{KOM09}
\par A second dimensionless ratio is acoustic scale 
$\ell_A(z)=(1+z)\frac{\pi d_A(z)}{d_s(z)}$.\cite{WAM07} 
CMB data determine $\ell_A(z_*)=302.10\pm0.86$.\cite{KOM09}
Comoving sound horizon $d_s$, in Hubble length units $c/H_0$,
is determined by
$r_s(z_*)=\int_0^{t_*}\frac{cdt}{a(t)\sqrt{3(1+R_b(t))}}$.
\par Fitting the modified Friedmann
equation to $H_0d_L(z)/c$ for $z\leq1$,\cite{MAN03,MAN06}
to shift parameter $R(z_*)$, and to acoustic scale ratio $\ell_A(z_*)$ 
determines model parameters $\alpha=0.7171, k=-0.01249,
\beta=0.3650\times 10^{-5}$. 
The fourth parameter is $\gamma=0.3976\times 10^{-8}$,
if fixed at $\gamma=3\beta/4R_b(t_0)$, which neglects dark matter.
There is no significant inconsistency with current empirical
data, given the demonstration by Mannheim\cite{MAN06} that type Ia
supernovae redshifts can be fitted neglecting 
$\Omega_m=-\beta/a^3H^2$ in the current epoch ($z\leq 1)$. Clearly   
parameter $\Omega_m$ must be reconsidered in the context of conformal
theory.
\par Parameters determined by the modified Friedmann equation fit
model-independent data from the recombination epoch $z_*=1090$ to
the present $z=0$.  Relevant computed parameters are
\begin{table}[h]
\begin{tabular}{lrrrrr}
$z             $&  1090&   100&    10&     1&     0\\
$H             $&  92.4& 11.50&  2.44&  1.43&  1.00\\
$\Omega_q      $& 0.474& 0.063& 0.625& 0.622& 0.271\\
$\Omega_\Lambda$& 0.000& 0.005& 0.121& 0.353& 0.717\\
$\Omega_k      $& 1.740& 0.963& 0.255& 0.025& 0.012\\
$\Omega_m      $&-0.555&-0.028&-0.001&-0.000&-0.000\\
$\Omega_r      $&-0.659&-0.003&-0.000&-0.000&-0.000\\
$\frac{H_0}{c}d_H    $& 15.54& 10.26&  4.71&  1.41&  1.00\\
$\frac{H_0}{c}d_\chi$&  614.1& 47.78&  5.68&  0.81&  0.00
\end{tabular}
\end{table}
\par $H(t)$ has its maximum value ${\bar H}=100.9$
for ${\bar z}=1371$, where $\Omega_q=1$.
As redshift $z$ increases, dark energy term $\Omega_\Lambda$ becomes
negligible.  For $z>{\bar z}$, $H(t)$ decreases to zero, while
$a(t)\to a_0=0.508\times 10^{-3}$ ($z_0=1967$) and $\Omega_q\to\infty$.
\par For negative $k$, as implied here, the geodesic Hubble radius 
becomes
$d_H=\sinh(\sqrt{-k}/aH)/\sqrt{-k}$, tabulated above as $H_0d_H/c$.
Similarly, geodesic distance to given redshift $z$ is
$d_\chi=\sinh(\sqrt{-k}\chi(z))/\sqrt{-k}$, where the comoving distance
is $\chi(z)=\int_{t_z}^{t_0}\frac{cdt}{a(t)}$.\cite{DOD03}
\par Meaningful cosmological structure requires the wavelength of a
periodic perturbation to be smaller than the Hubble horizon,\cite{AWB08}
so that for comoving wavevector magnitude $\kappa$, $\kappa d_H>1$.
For angular structure, multipole index
$\ell\simeq\kappa d_\chi=\kappa d_H\frac{d_\chi}{d_H}$.\cite{AWB08}
The relative scale criterion is $\ell>\frac{d_\chi}{d_H}$.
Using parameter values listed above, conformal theory
implies $\ell(z_*)>39.52$, consistent with observed CMB 
anisotropies.\cite{DOD03}
\par Defining criterion $\zeta=\sxth R-w^2$, the dimensionless sum
rule determines $\zeta=\xi_0+\xi_1-w^2=
H(t)^2(2\Omega_q+\Omega_m+\Omega_r)$.  For $a\to 0$, when both $\alpha$
and $k$ can be neglected, the sum rule implies
$\zeta=H(t)^2(\Omega_q+1)$.  For large $a$, $\zeta=H(t)^2(2\Omega_q)$.
$\zeta>0$ in both limits, regardless of numerical values, since
$\Omega_q>0$.  Using the present empirical parameters, $\zeta$ is
positive for all $z$.  For nonzero $\phi_0^2$, this implies that
empirical scalar field parameter $\lambda$ is negative.
\par To test consistency, parameter $H_0d_V(z)/c$, where
$d_V(z)=[d^2_\chi(z)cz/H(z)]^\thrd$,\cite{EIS05} was computed for  
$z=0.35$.  For $\alpha=0.732$ and vanishing $k,\beta,\gamma$, computed 
value 0.3087 agrees with empirical 0.322$\pm$0.015.\cite{EIS05} 
Using all four adjusted parameters, the computed value 
changes only to 0.3083.

\section{Conclusions and speculations}
\par Conformal theory provides a straightforward explanation of
dark energy: it appears in the energy-momentum tensor of
the scalar field required by the Higgs mechanism to produce the masses
of gauge bosons. 
It should be possible to compute the implied cosmological constant 
as the self-interaction of the Higgs scalar field.\cite{NES08}
The required transition amplitude, which creates an induced gauge field,
depends on the extremely small cosmological time derivative
of the dressed scalar field.  The time constant is of the
order $10^{10}$ years.
Solving the coupled field equations for $g_{\mu\nu},\Phi$, and induced
U(1) gauge field $B_\mu$, using this computed time derivative, gives
$w\simeq2.651\hbar H_0=3.984\times 10^{-33}eV$.\cite{NES10a}  
This approximate calculation agrees in order of magnitude with 
parameter value $w=1.273\times 10^{-33}eV$ implied by present value
$\Omega_\Lambda(t_0)=\alpha=0.7171$.
Details will be published separately.\cite{NES10a}
These numbers justify the conclusion that conformal theory
explains both the existence and magnitude of dark energy.
\par The modified Friedmann equation predicts a stationary value of
the Hubble function at redshift somewhat greater than $z_*$,
preceded by an inflationary epoch. A time-dependent theory of the 
relevant field parameters is needed in order to compare with current 
big-bang theory.
\par Dark matter is not required for $z\leq 1$ supernovae redshifts, 
for anomalous galactic rotation,\cite{MAN06} or for the present 
empirical parametrization of the modified Friedmann equation.
Interpretation of the parameter $\Omega_m$ requires substantial 
revision if the modified Friedmann equation is correct.
\par Models of gravitational lensing should consider that geodesic
deflection is due to the difference between nonuniform galactic matter
and the nonvanishing averaged background that determines the RW metric. 
This subtracts the isotropically dispersed repulsive field considered
here from the attractive field due to the observed galaxy.  
Subtracting this isotropic repulsive field would have the
same effect as an additive attraction attributed to a dark-matter halo. 

\end{document}